\documentclass[12pt]{iopart}

\usepackage{graphicx}

\begin{document}

\title{Evaluating links through spectral decomposition}

\author{Gonzalo Travieso and Luciano da Fontoura Costa} 
\address{Instituto de F\'{\i}sica de S\~ao Carlos. 
Universidade de S\~ ao Paulo, S\~{a}o Carlos, SP\\
PO Box 369, 13560-970, Brazil\\
\eads{\mailto{gonzalo@ifsc.usp.br}, \mailto{luciano@ifsc.usp.br}}}

\begin{abstract}   
  Spectral decomposition has been rarely used to investigate complex
  networks.  In this work we apply this concept in order to define two
  types of link-directed attacks while quantifying their respective
  effects on the topology.  Several other types of more traditional
  attacks are also adopted and compared.  These attacks had
  substantially diverse effects, depending on each specific network
  (models and real-world structures).  It is also showed that the
  spectral-based attacks have special effect in affecting the
  transitivity of the networks.
\end{abstract}


\maketitle

The theory of graphs (e.g.~\cite{Tutte:2001}) and networks
(e.g.~\cite{Albert_Barab:2002,Newman:2003}) represents one of the most
multidisciplinary, integrated and applicable areas of theoretical
mathematics and computing.  Although its origin is often traced back
to Euler's solution of the K\"onigsberg bridge problem, graphs have
been around for much longer, at least since the first map was draw on
sand~\footnote{Maps are a special kind of graph called
  \emph{geographical}, which is characterized by the fact that the
  edges have well-defined positions in an embedding space.}.  Because
graphs and networks can represent most discrete structures possibly
underlying dynamical systems~\cite{Newman:2003,Boccaletti_etal:2006},
they are particularly useful for modeling a vast range of problems.
The identification of structured connections in growing graphs,
especially the existence of hubs and their fundamental importance
(e.g.~\cite{Albert_Barab:2002}), helped to catalyse a surge of
interest which had already been sparkled by random graphs and small
world networks studies, giving rise to the new theory of \emph{complex
  networks}.

A good deal of the investigations in complex networks have focused on
relatively simple properties such as the node degree (i.e.\ the number
of connections established by a node), clustering coefficient (i.e.\
the degree of interconnectivity among the immediate neighbors of a
node) and the shortest path length between two nodes.  These
measurements÷\cite{Costa_surv:2006} are particularly important because
they correspond to the distinguishing features of the main complex
network models.  For instance, small world networks are characterized
by low mean shortest path length together with high clustering
coefficient, and scale free networks exhibit power law degree
distributions.  However, as these measurements are not enough to
provide a complete, invertible, representation of the complex network
of interest, they will not be enough to directly express many
important connectivity properties.

There are so many possible measurements of complex networks that it is
useful to organize them into categories (e.g.~\cite{Costa_surv:2006}).
A particular interesting and useful category of measurements are those
called spectral (e.g.~\cite{Biggs:1993,Chung:1997, Cvetkovic:1997}),
in the sense of involving the eigenvalues of the adjacency matrices of
the analyzed graphs.  Spectral approaches to graphs and networks are
particularly important because of many reasons including the
relationship between eigenvalues and the dynamics of the network,
connectedness, cuts, modularity and cycles, among others. Such
concepts and methods have progressively attracted the attention from
the complex networks community, to the extent that some of the best
community finding algorithms in this area are now based on spectral
methods (e.g.~\cite{Newman:2006}).

Spectral approaches in graphs, have many relationships with
theoretical and applied physics, they may consider the adjacency
(e.g.~\cite{Biggs:1993,Cvetkovic:1997}) or Laplacian matrices
(e.g.~\cite{Chung:1997}) of graphs.  In this work we concentrate
attention in the former type of approaches.  More specifically,
because the spectrum of a graph does not provide a complete
representation, we focus our attention on the possibility to use the
\emph{eigenspaces} of graphs~\cite{Cvetkovic:1997} in order to derived
more powerful features for characterizing the graph connectivity.
Complete representations are important in graph studies because they
allow a one to one mapping between any graph (including its
isomorphisms) into a feature space which can be used for unambiguous
graph classification (e.g.~\cite{Costa_surv:2006}), avoiding
\emph{degenerate} mappings~\footnote{In a degenerate mapping, two or
  more different graphs can be mapped into the same representation,
  precluding the map inversion.  Degenerate mappings are frequently
  used for network~\emph{characterization} and classification.}. While
any graph can be precise and completely represented in terms of its
spectrum and eigenspaces, such formulation ultimately depends on the
node labeling for the correct identification of the eigenspaces.
Therefore, such a representation is not invariant to node label
permutations and graph isomorphisms.  While the existence of a
complete and invariant representation of graphs does not seem to be
likely (e.g.~\cite{Cvetkovic:1997}), it is still interesting to
consider additional features rather than just the graph spectrum.  One
of the most natural such a complementation can be achieved by
considering also the eigenspaces of the graphs.

We consider here the problem of quantifying the importance of links in
networks using the spectral decomposition of the adjacency matrix.
Based on link spectral measurements that are described below, a
fraction of the links is removed and the effect of this removal on
network topology is quantified for some specific model or real
networks.  For comparison's sake, the same procedure is also applied
using other, non-spectral, link measurements.  There are many works
dealing with vulnerability of model and real networks to attacks on
nodes or links
\cite{latora05:_vulner,holme02:_attac,He20092243,martin06:_random,motter02:_range,wang08:_robus,gao06:_between}.
Those works do not use spectral measurements.  Spectral techniques are
often used to express centrality measures of nodes
\cite{estrada08:_graph} and for community detection
\cite{Newman:2006,ma10:_eigen,qin09:_commun_findin_scale_free_networ},
but were also used for the network vulnerability problem
\cite{yazdani10,green09:_small_scotl,jamakovic07}.  None of these
works used spectral decomposition.  Spectral decomposition is used in
Ref.~\cite{PhysRevE.81.016101}, where the authors consider the problem
of reconstructing a network after perturbation.

This article is organized as follows.  First, the basic concepts from
complex network (e.g.~\cite{Albert_Barab:2002,Newman:2003,
  Boccaletti_etal:2006, Costa_surv:2006}) and eingenspace
(e.g.~\cite{Biggs:1993,Cvetkovic:1997}) theories are presented in an
introductory and self-contained way.  Then the experimental
methodology is explained regarding the generation of the synthetic
complex network models and measurements used for the evaluation, which
is followed by the presentation and discussion of the results.

\section{Basic Concepts}

A \emph{graph} (or \emph{complex network}) $\Gamma=(V,E)$ is a
discrete structure composed by a set of vertices or nodes $V$ and a
set of edges or links $E$, with $N=|V|$ and $L=|E|$~\footnote{The
  operation $|X|$ stands for the cardinality of the set $X$, i.e. its
  number of elements.}.  We henceforth assume that the nodes of such a
graph are labelled with successive positive integer values, i.e. $1,
2, \ldots, N$, and that multiple or self-connections are not present.
The existence of an edge extending from node $i$ to node $j$ is
indicated, in the case of undirected graphs considered here, by the
unordered pair $(i,j)$.  A graph can be completely specified in terms
of its \emph{adjacency matrix} $\mathbf{A}$ of dimension $N \times N$.
The presence of the edge $(i,j)$ is indicated as $A_{ij}=A_{ji}=1$;
otherwise $A_{ij}=A_{ji}=0$.  Note that the trace of $\mathbf{A}$ is
zero. A graph is said to be \emph{connected} in case any node can be
reached from any other node while moving through the edges of the
graph.  The sequence of nodes (or edges) visited during such movements
is called a \emph{path} of the graph, with length equal to the number
of involved edges.

The \emph{characteristic polynomial} of the adjacency matrix
$\mathbf{A}$ is given as $P_{\Gamma}(\lambda) = \mathrm{det}(\lambda
\mathbf{I} - \mathbf{A})$.  The eigenvalues of $\mathbf{A}$, assumed
to be sorted and represented as $\lambda_1 \geq \lambda_2 \geq \ldots
\geq \lambda_N$, correspond to the zeros of $P_{\Gamma}(\lambda)$.
Because the graph is undirected, $\mathbf{A}$ is symmetric, implying
real eigenvalues.  Each of these eigenvalues is a solution of the
equation $\mathbf{A} \mathbf{v}_i = \lambda_i \mathbf{v}_i$ for a
non-zero vector $\mathbf{v}_i$ called the \emph{eigenvector}
associated with $\lambda_i$.  The set of the $N$ eigenvalues is called
the \emph{spectrum} of $\Gamma$.  The largest eigenvalue $\lambda_1$
of a graph $\Gamma$ is the \emph{index} of $\Gamma$.  In a connected
graph, the eigenvector associated to the index has all its elements
positive and is called the \emph{principal eigenvector} of $\Gamma$.

Because the adjacency matrices obtained by changes of the labels of
the nodes are similar (and the graphs isomorphic), the spectrum of a
graph does not depend on the node labeling and is therefore
an~\emph{invariant} to label permutations.  Note that, unlike
eigenvalues, eigenvectors are not invariant to label permutations.

The original adjacency matrix can be expressed in terms of its
\emph{spectral decomposition} given as
\begin{equation}  
  \mathbf{A} = \sum_{i=1}^{N} \lambda_i \mathbf{S}_i ,
  \label{eq:spec_dec}
\end{equation}
where $\mathbf{S}_i = \mathbf{v}_i \mathbf{v}_i^T$, $i = 1, 2, \ldots,
N$.

It follows immediately from Equation~(\ref{eq:spec_dec}) that a graph
can be completely specified in terms of its spectrum and eigenvectors.

\section{Evaluation methodology}

\paragraph{General insight}

We use the entry in line $i$, column $j$ of some components of the
spectral decomposition of the adjacency matrix as an indication of the
``importance'' of the link between nodes $i$ and $j$.  If all
components are included, the adjacency matrix is recovered, and the
importance is equal to $1$ for all the links.  We need to choose some
components, with two possibilities presenting themselves: specifying a
fixed number of the most important eigencomponents or specifying a
threshold, including all components with eigenvalue above the
threshold.  We consider a simple case of each of these possibilities:
\begin{itemize}
\item The component corresponding to the largest eigenvalue, here
  called the \emph{largest eigencomponent}, that is, the matrix given
  by
  \begin{equation}
    \label{eq:largeigen}
    \lambda_1 \mathbf{S}_1
  \end{equation}
  as in Equation~\Eref{eq:spec_dec}.
\item The sum of the contribution of all positive eigenvalues in
  Equation~\Eref{eq:spec_dec}, here called \emph{positive
    eigencomponent}:
  \begin{equation}
    \label{eq:poseigen}
    \sum_{i, \lambda_i>0} \lambda_i \mathbf{S}_i.
  \end{equation}
\end{itemize}

\paragraph{Evaluation}

To compare different ``importance'' measurements we compute them for
the same network and subsequently attack a fraction of the most
important links in the network according to each measurement.  The
link ``importance'' measurement whose attack yields the most
significant impact on the network structure (as evaluated by some
network measurements) is considered the most effective.

\paragraph{Networks used}

We used three network models and two real-world networks.  The models
used are:
\begin{description}
\item[Erd\H{o}s-R\'enyi (ER)] Used as a null model.
\item[Barab\'asi-Albert (BA)] Chosen because of the presence of hubs.
\item[Watts-Strogatz (WS)] A network model with high clustering.
\item[Holme-Kim (HK)] The growing scale-free network model with finite
  transitivity proposed by Holme and Kim \cite{HolmeKim:2002}.
\end{description}

The real-world networks used are the US Airports and the US Power Grid
networks, both from the Pajek datasets~\cite{pajekdata}.  These
networks were chosen because the attacks to their links have a clear
meaning.

\paragraph{Measurements}

We use two kinds of measurements: the ones used to rank the links and
the ones used to evaluate network topology.

\subparagraph{Link ranking}

We propose that the ``importance'' of link $(i,j)$ be taken as the
entry in row $i$, column $j$ of the matrices defined by
Equations~\Eref{eq:largeigen} and \Eref{eq:poseigen}.

For comparison, we evaluate also the following possibilities:
\begin{description}
\item[Betweenness centrality] of the link.
\item[Degree product] corresponds to the product of the degrees of the
  nodes at both ends of the link.  This measurement was also used in
  Ref.~\cite{kaiser04:_edge}.
\item[Random walk] centrality, computed as the fraction of steps of
  random walkers started at random nodes that pass through the link.
  We used $10 N$ random walkers ($N$ is the number of nodes in the
  network), each one with $100$ steps.
\item[Number of triangles] The number of triangles [sets of links of
  the kind $(i,j), (j, k), (k,i)$ including the link.
\item[Random] order, that is, nodes are attacked at random.
\end{description}

\subparagraph{Network topology}

To evaluate the network topology, we consider the following
measurements:
\begin{description}
\item[Transitivity] (sometimes called also clustering coefficient) is
  a simple example of a measurement of local network connectivity.
\item[Average path length] This measurement, also called distance,
  help quantify how the accessibility of nodes is being affected by
  the attack.
\item[Largest degree] Looking at the largest degree in the network we
  can evaluate the effect of the attack in the most important hubs.
\item[Number of clusters] As the links are removed, the network breaks
  in independent clusters.  The number of cluster gives a measure of
  the fractioning of the network.
\item[Largest cluster] The size of the largest cluster, measured as
  the fraction of nodes in this cluster.
\item[Number of squares] The number of squares [sets of links of the
  kind $(i,j), (j,k), (k,l), (l,i)$, for distinct nodes $i,j,k,l$]
  including the link.
\end{description}

\subparagraph{Attack}

An attack using a given link measurement is simulated by the following
procedure: First the measurement is computed for each link; the links
are then ranked from largest to smallest value, and a given fraction
of the links with the largest values is removed from the network.
Finally, network measurements are computed for the attacked network.

\section{Results and Discussion}

The experiments were run using networks of $1000$ nodes and average
degree $4$.  For Watts-Strogatz networks a rewiring probability of
$0.05$ was used; for the Holme-Kim networks, the probability of a
triad formation step is $0.8$.  The results are averages of $50$
networks for each model.

\begin{figure*}
\includegraphics{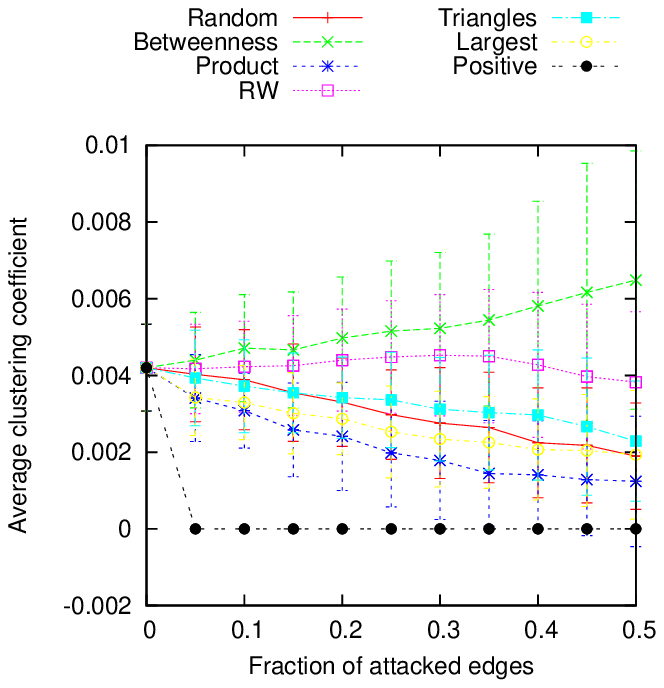}
\includegraphics{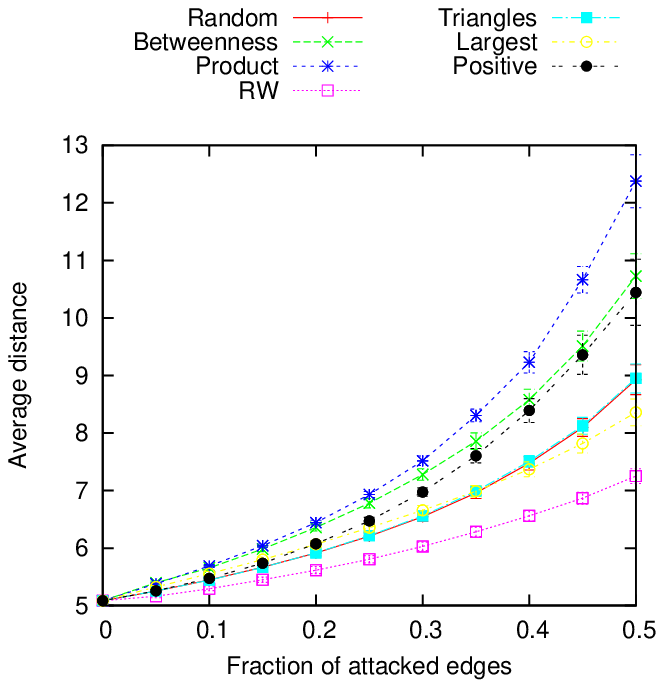}\\
\includegraphics{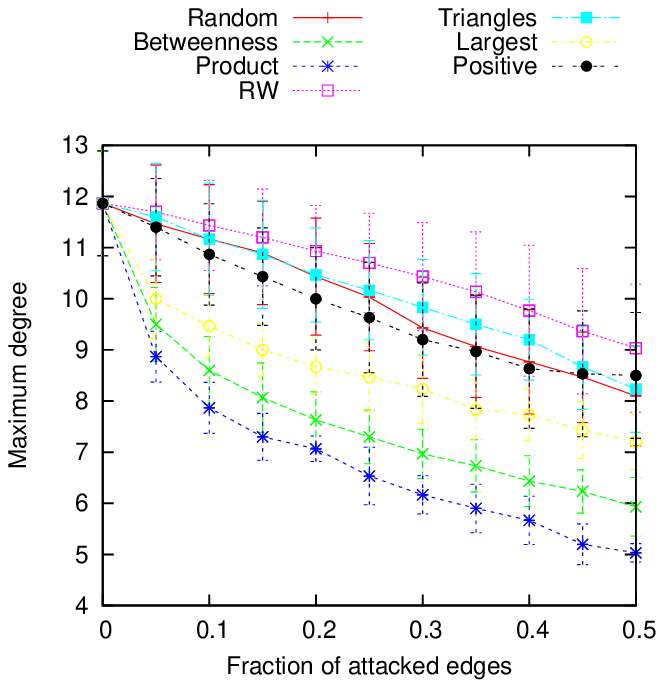}
\includegraphics{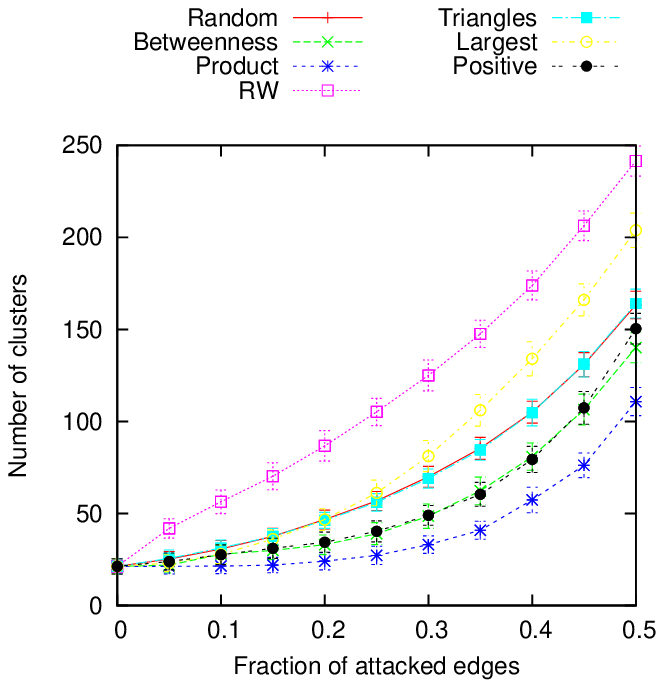}\\
\includegraphics{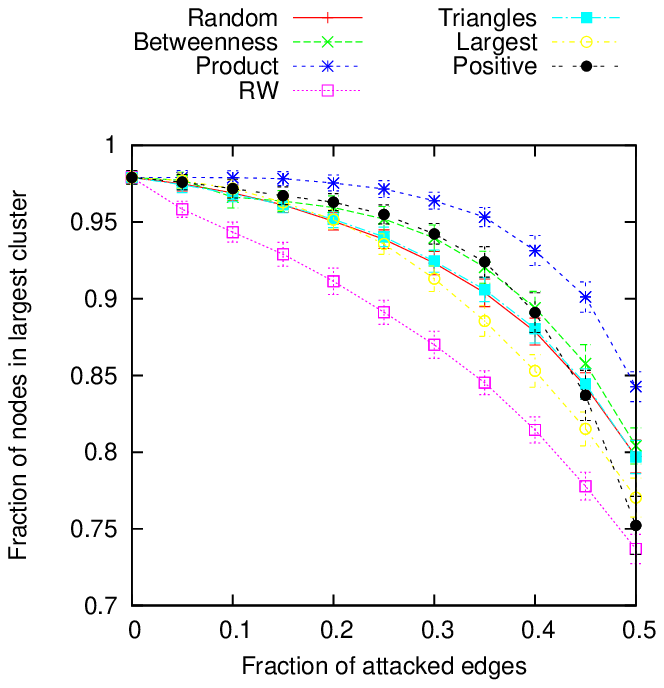}
\includegraphics{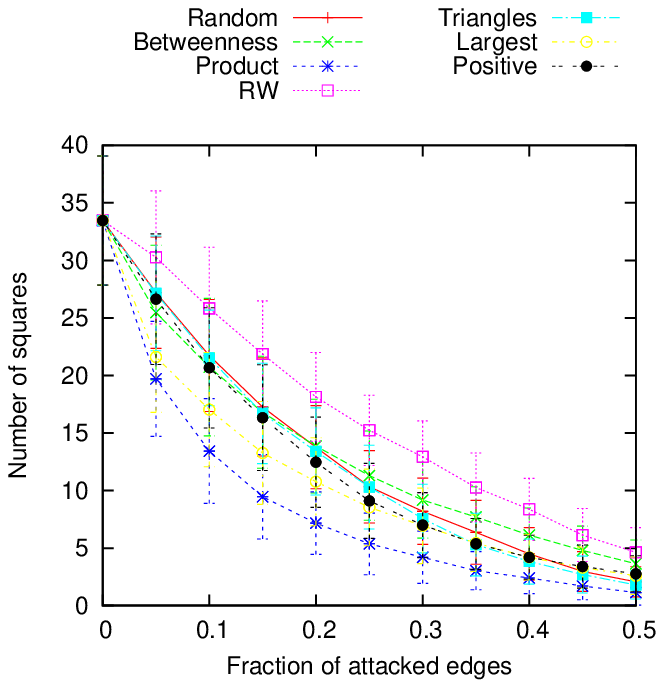}\\
\caption{Effect of link attacks to some network measurements for
  Erd\H{o}s networks.}\label{fig:attackerdos}
\end{figure*}

\begin{figure*}
\includegraphics{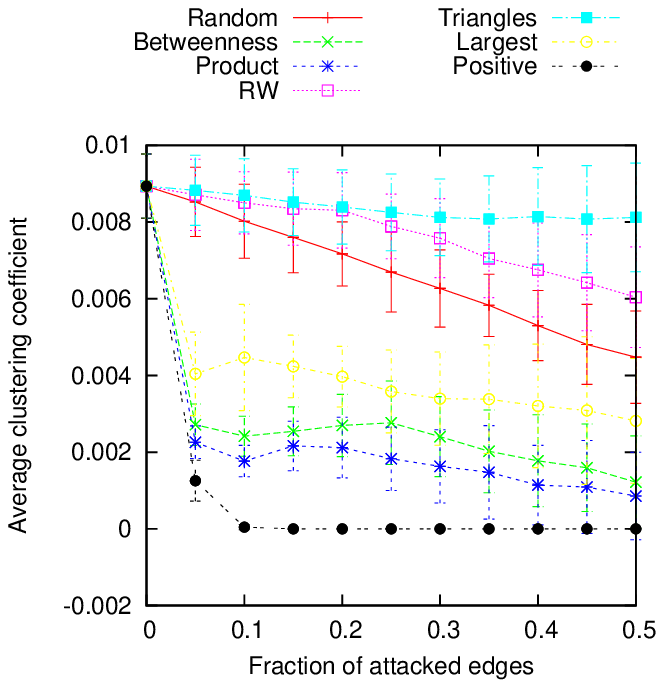}
\includegraphics{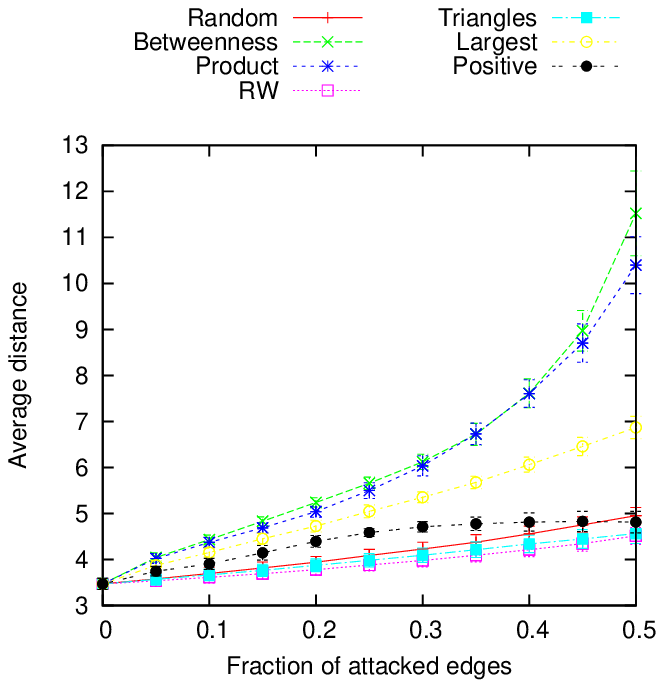}\\
\includegraphics{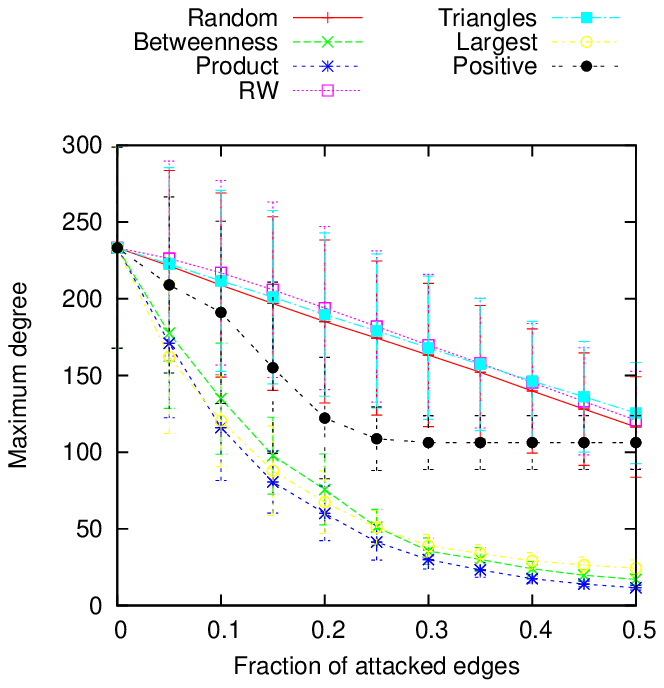}
\includegraphics{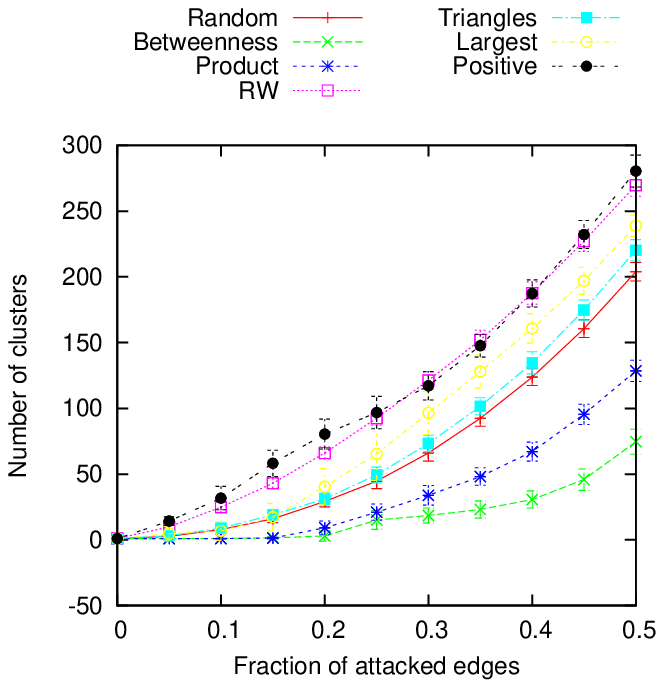}\\
\includegraphics{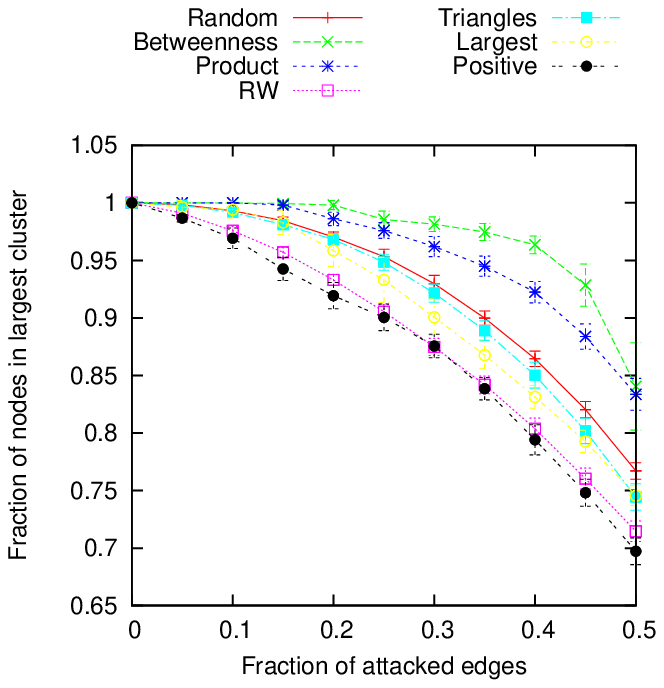}
\includegraphics{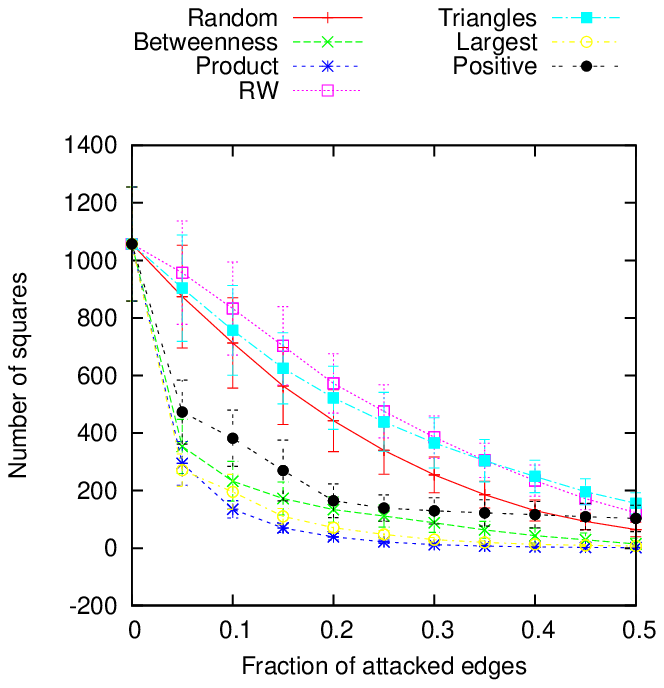}\\
\caption{Effect of link attacks to some network measurements for
  Barab\'asi networks.}\label{fig:attackbarabasi}
\end{figure*}

\begin{figure*}
\includegraphics{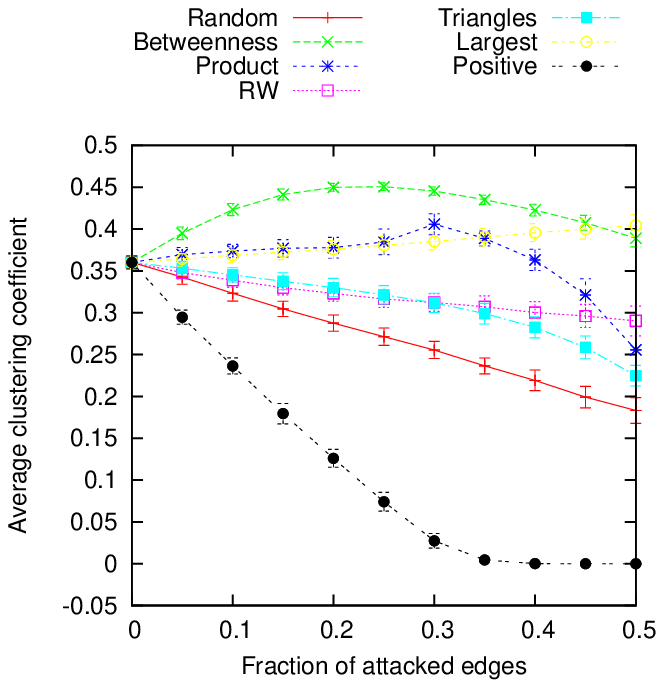}
\includegraphics{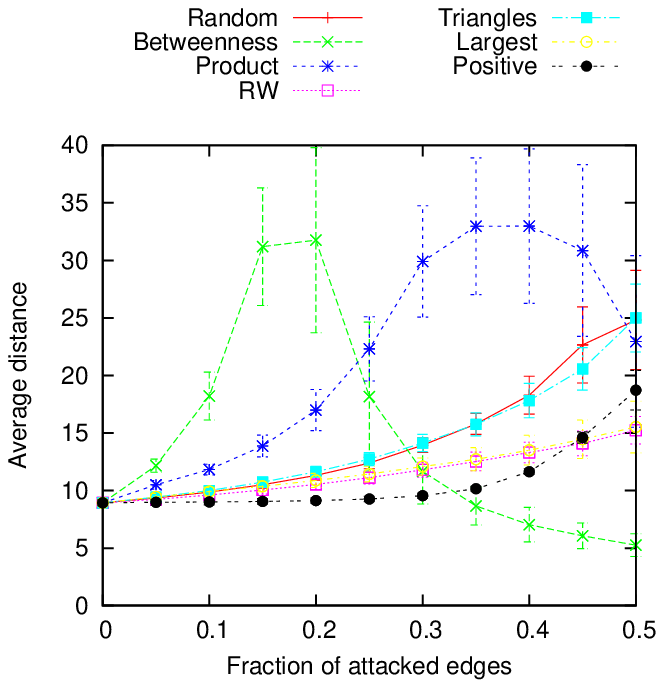}\\
\includegraphics{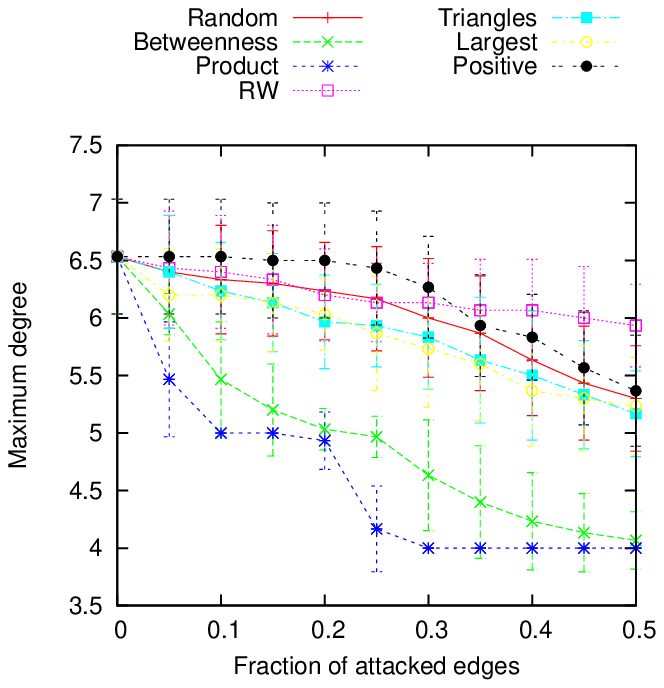}
\includegraphics{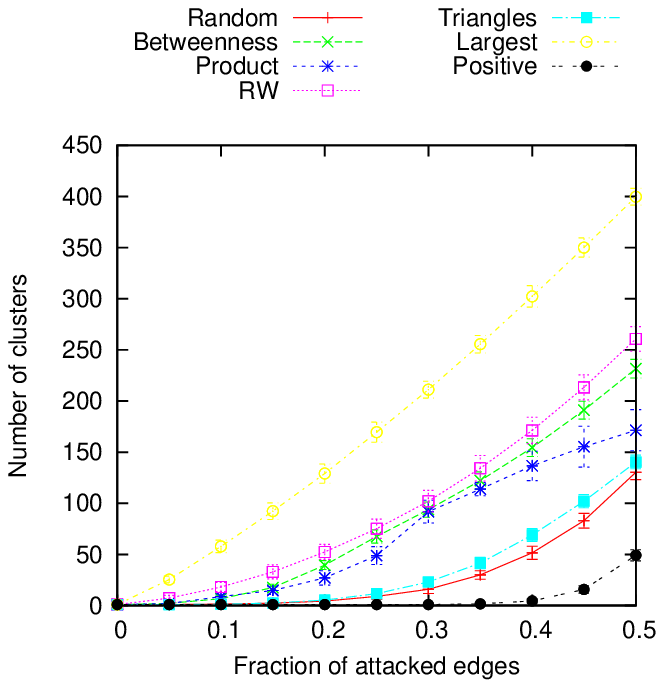}\\
\includegraphics{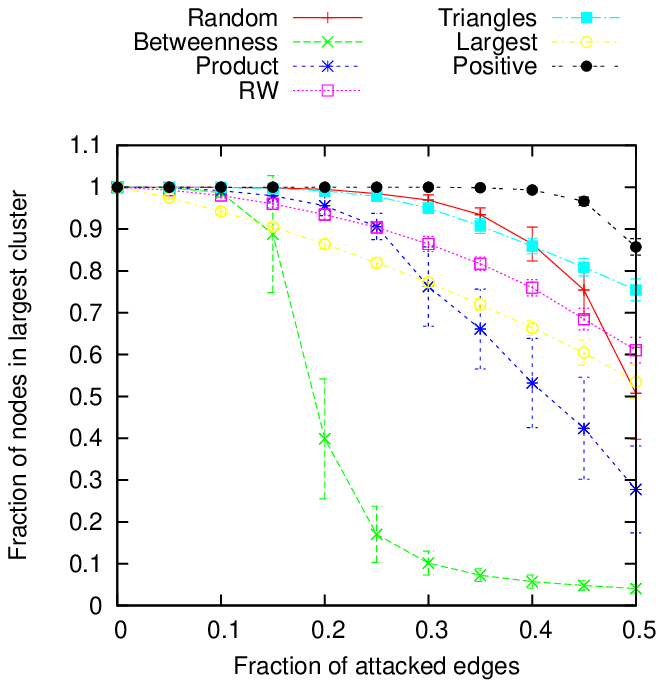}
\includegraphics{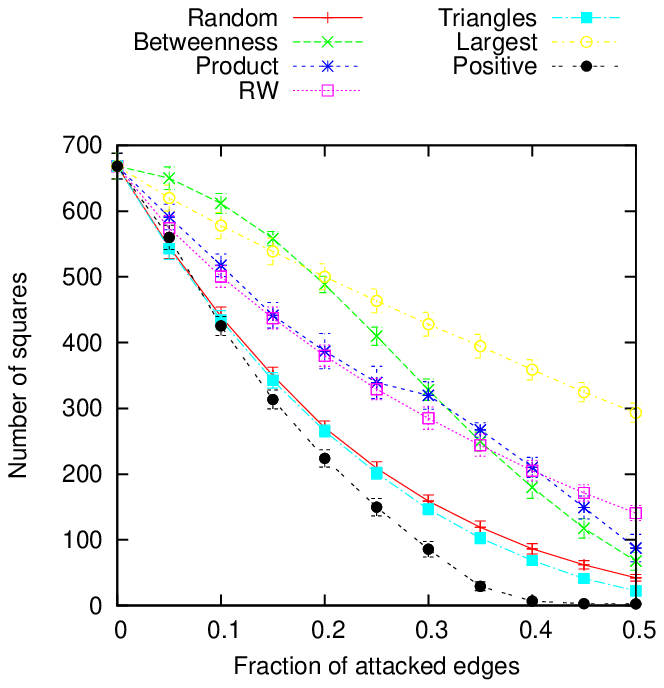}\\
\caption{Effect of link attacks to some network measurements for
  Watts-Strogatz networks.}\label{fig:attackwatts}
\end{figure*}

\begin{figure*}
\includegraphics{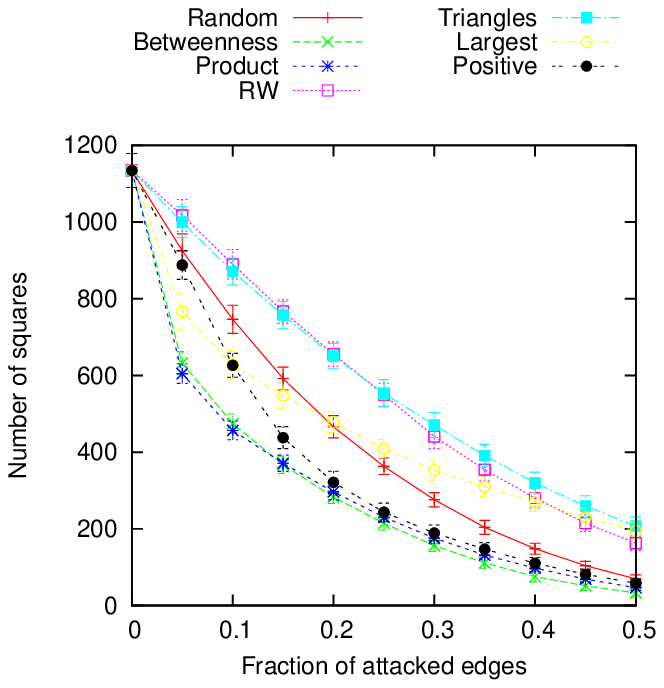}
\includegraphics{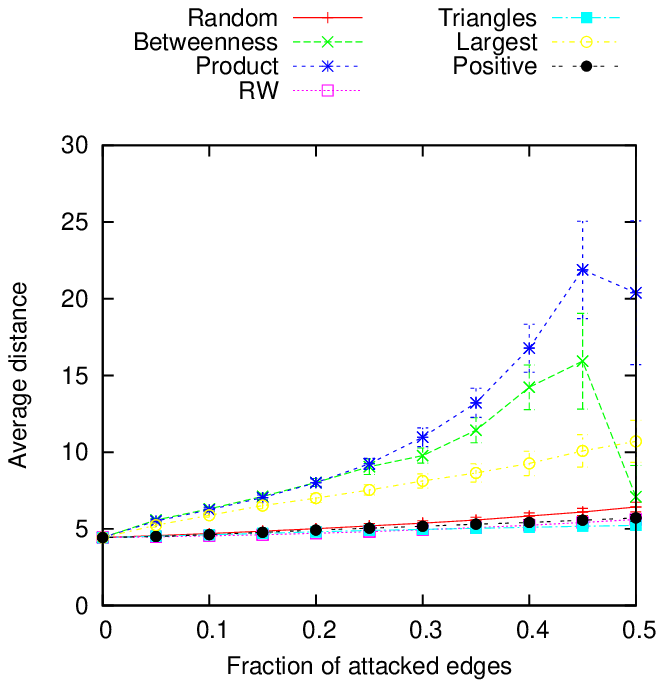}\\
\includegraphics{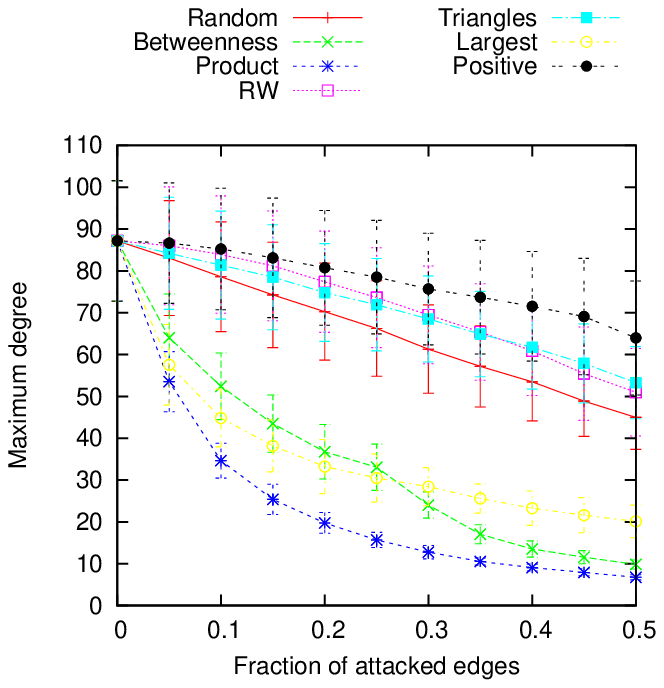}
\includegraphics{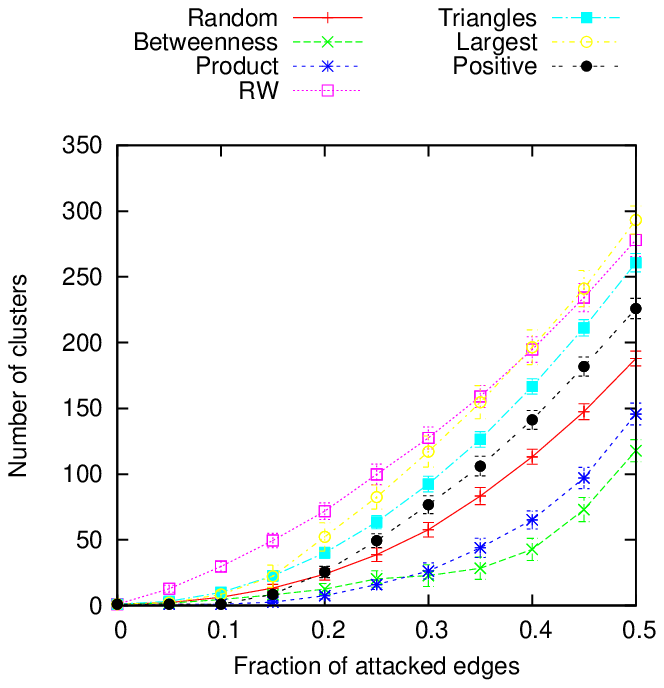}\\
\includegraphics{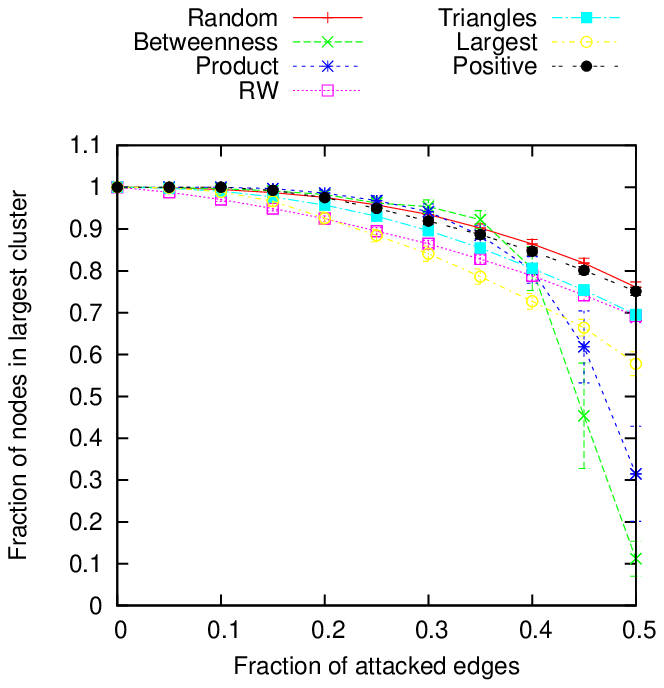}
\includegraphics{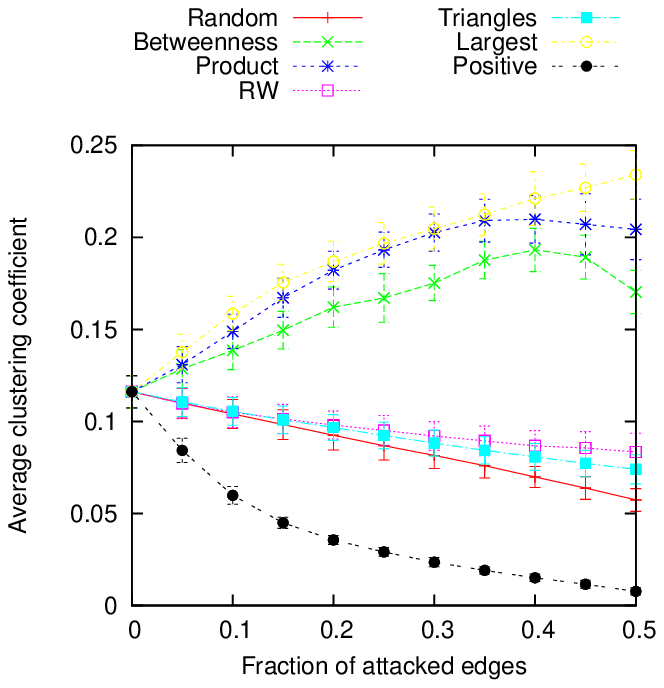}\\
\caption{Effect of link attacks to some network measurements for
  Holme-Kim networks.}\label{fig:attackholme}
\end{figure*}

\begin{figure*}
\includegraphics{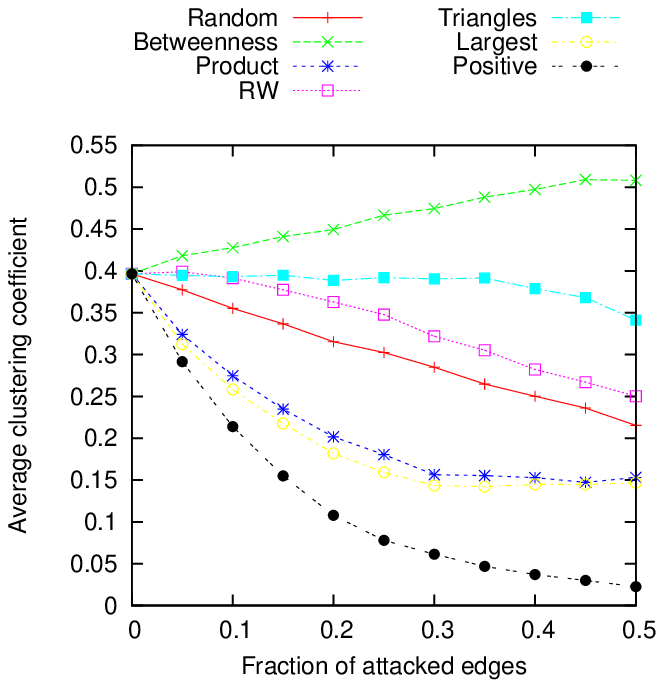}
\includegraphics{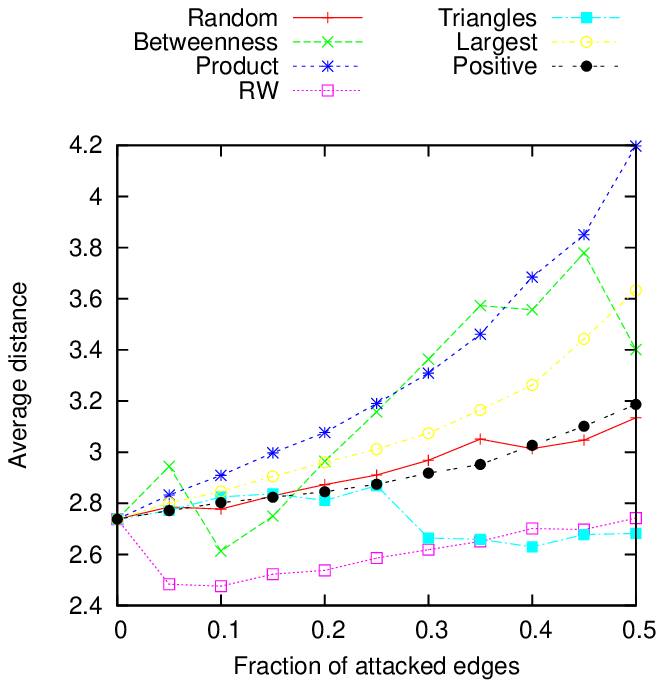}\\
\includegraphics{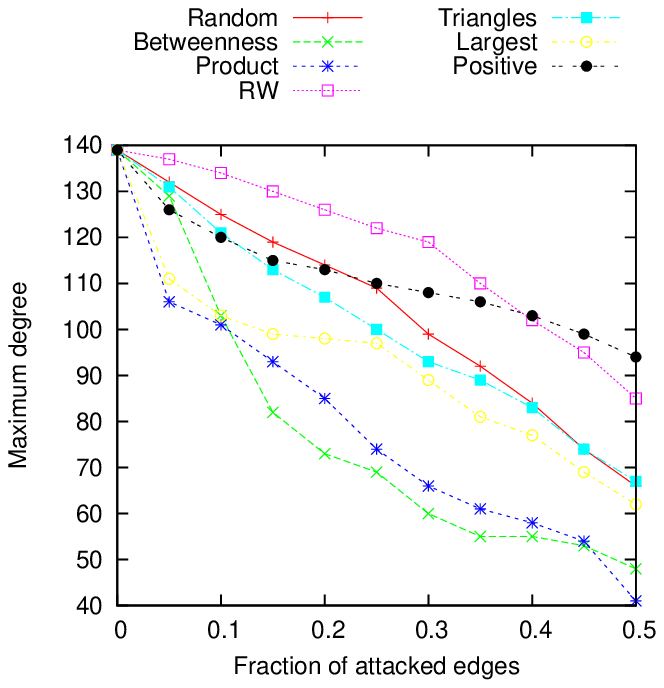}
\includegraphics{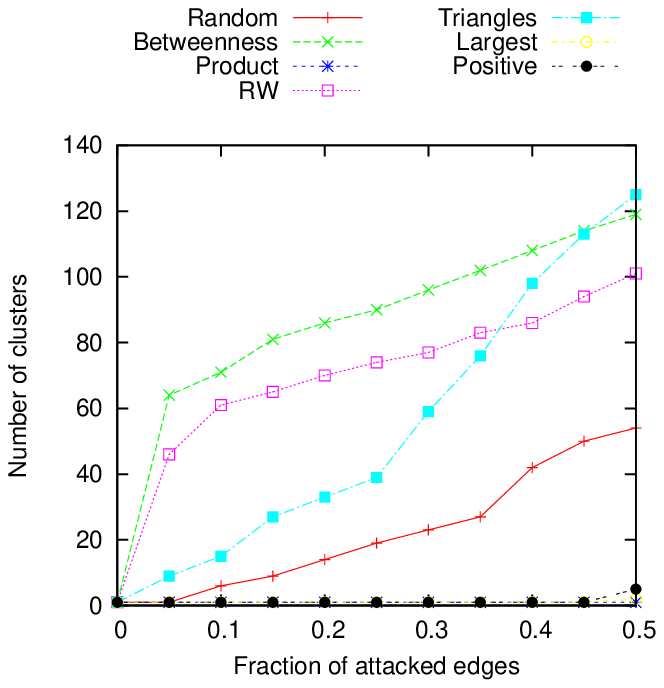}\\
\includegraphics{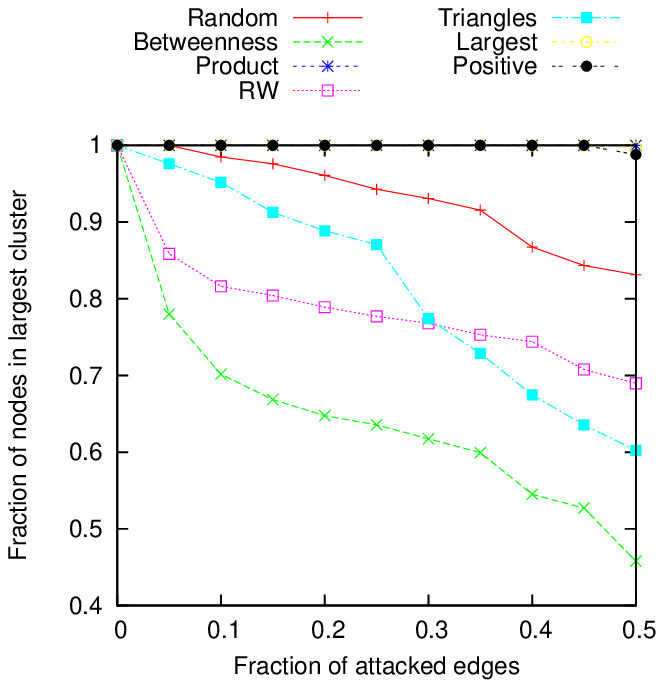}
\includegraphics{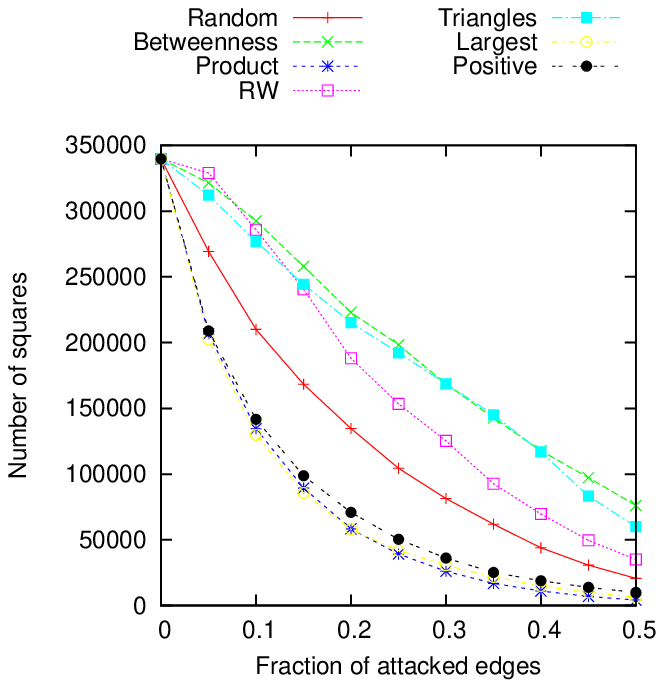}\\
\caption{Effect of link attacks to some network measurements for the
  US Airport network.}\label{fig:attackair}
\end{figure*}

\begin{figure*}
\includegraphics{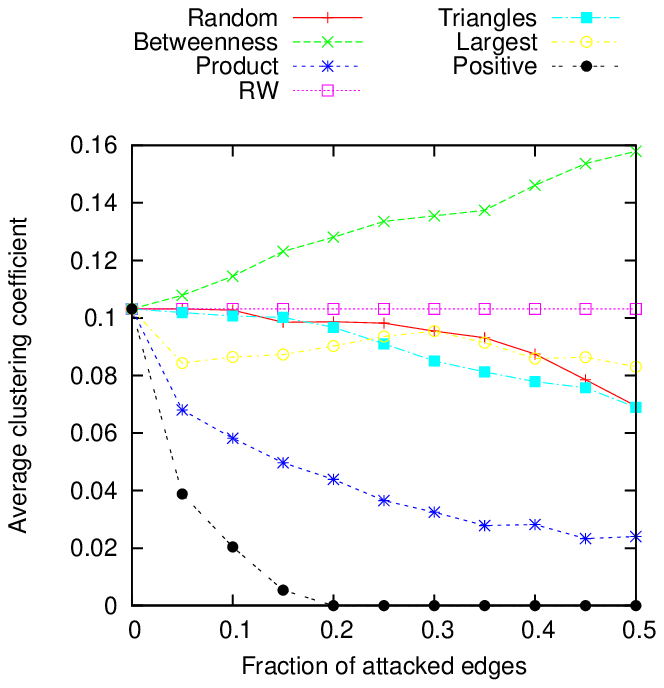}
\includegraphics{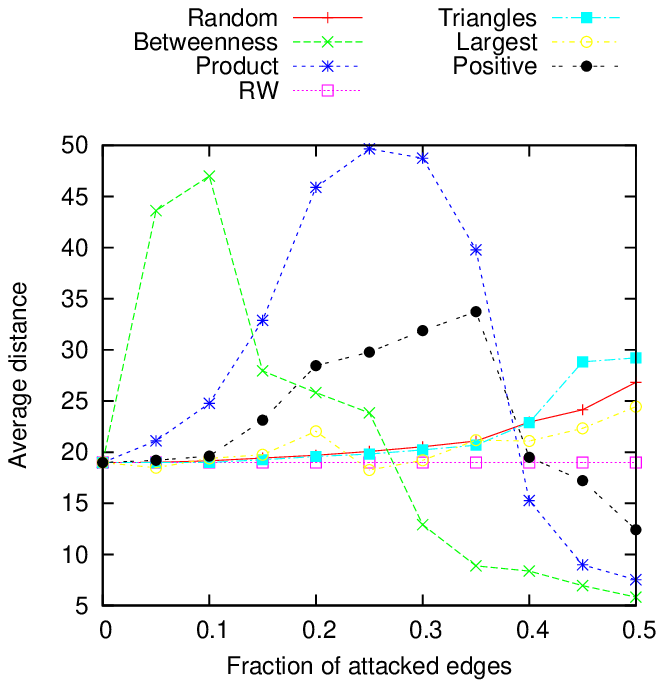}\\
\includegraphics{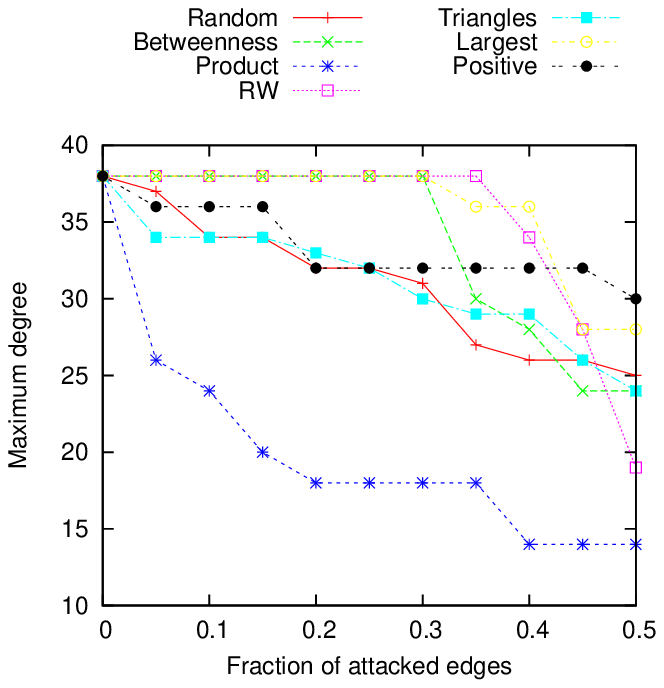}
\includegraphics{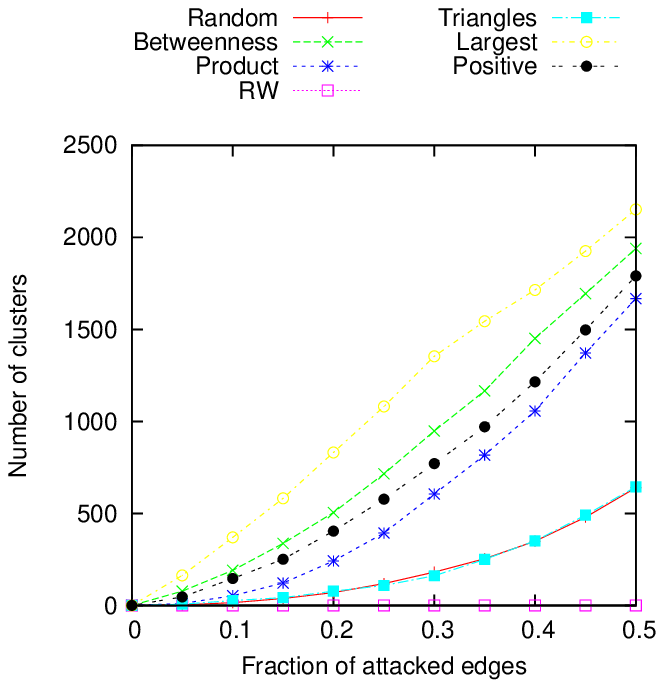}\\
\includegraphics{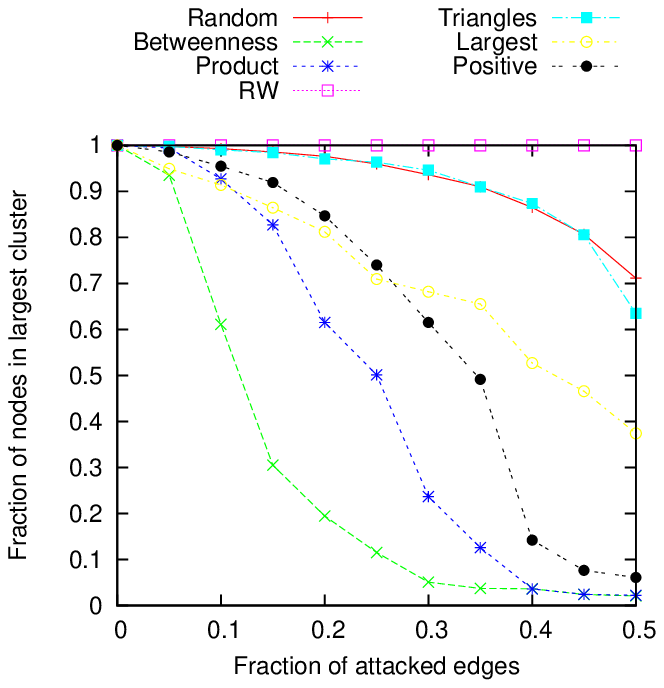}
\includegraphics{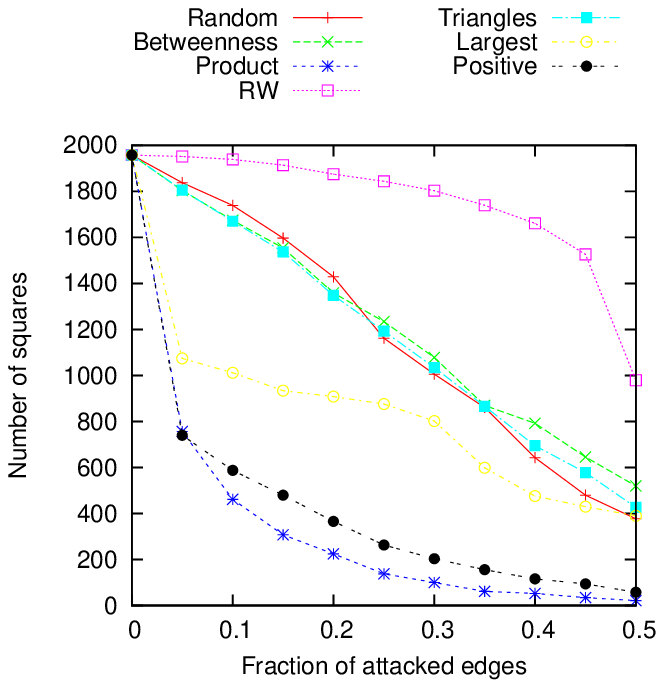}\\
\caption{Effect of link attacks to some network measurements for the
  US Power Grid network.}\label{fig:attackpower}
\end{figure*}

The results for model networks are shown in
Figures~\ref{fig:attackerdos} (ER), \ref{fig:attackbarabasi} (BA),
\ref{fig:attackwatts} (WS), and \ref{fig:attackholme} (HK).  For real
networks, the results are shown in Figures~\ref{fig:attackair} (US
Airports) and \ref{fig:attackpower} (US Power grid).

Attacks directed by triangle count have mostly similar results as
random attacks, with the exception of the HK model and the US Airports
network, where they tend to break the network clusters and preserve
the squares.  The maximum degree is most affected by attacks following
the degree product, but also strongly affected by the largest
eigencomponent and betweenness.

Attacks by degree product have also strong effect in the average
distances, but other attacks have different effects on different
topologies: for ER networks, betweenness and positive eigencomponents
have similar and important effects; for BA, betweenness has almost the
same effect than degree product and they are followed by the largest
eigencomponent, with other attacks fairing similarly to random
attacks.  For HK networks, the results are similar to that for the BA
networks, but betweenness is less effective.

Without considering the ER and BA model, that have too small
transitivity, we see that the transitivity is reduced fast by attacks
based on positive eigencomponents. The effect of attacks based on
random walk betweenness is not much different from random attacks.
The other attacks, specially by betweenness, increase the
transitivity, i.e.\ they tend to preserver triangles while edges are
removed.

The strongest effect on the clustering structure of the network is
achieved by attacks based on random walk betweenness or largest
eigencomponent (for number of clusters) and random walk betweenness
and link betweenness (for size of the largest cluster).  For BA
networks the positive eigencomponents attacks are as effective as
random walks, while for HK networks this happens with the largest
eigencomponent attacks.  Other attacks are similar or less effective
than random attacks, but note how in attacks by product degreee, the
positive and largest eigencomponents tend to preserve the clustering
structure of the airports network.

Attacks by degree product are the most effective to destroy squares in
the studied networks. On the other hand, attacks by random walk
betweenness tend to preserve links that take belong to squares.  Link
removal by betweenness centrality tends to increase the average
distances, while preserving the transitivity.  Attacks guided by the
largest eigencomponent are specially effective to dismember the
network in a large number of clusters (with the exception of the
airports network).  When guided by the positive eigencomponent, the
attacks clearly destroy the transitivity of the network, but their
effect on other measurements is mixed and topology dependent.  Degree
product can also be used in attacks to increase the average distances,
and reduce the transitivity.  However their effect on other
measurements, although distinct, changes from one type of network to
another.

There is also a curious behavior for the power grid network: for the
number of cluster, the largest eigencomponent is the most effective
type of attack, while it has smaller effect on the size of the largest
cluster.

\section{Concluding Remarks}

Despite the increasing attention on the effects of complex network
structure on vulnerability to attacks, relatively little attention has
been paid to spectral approaches.  In this work we reported an
investigation where spectral decomposition of the adjacency matrix is
applied in order to direct the attacks, while the respective effects
on the network topology are identified.  More specifically, we used
the eigencomponent associated to the highest eigenvalue and the sum of
all positive eigenvalues.  Several interesting results were
identified.  The most definite verified effect is the sharp decrease
of the transitivity of the network implied by attacks guided by the
largest eigencomponent.  This phenomenon has been found to be even
stronger than attacks oriented to triangles.  The effect of the
diverse types of attacks on the network topology depended largely on
the specific network types.  For instance, the attacks by positive
eigencomponents had no effect in changing the number of clusters in
the US Airport network, while it implies a marked increase in the
number of clusters in the case of the US Power Grid network.  In
addition, the attacks by largest eigencomponents yielded the sharpest
increase of number of clusters in the WS model, while having less
strong effects in the other networks.  Though spectral decomposition
approach has seldom been used in complex networks research, our
results show that it is capable of emphasizing specific aspects of the
network topology.  It is therefore expected that additional
applications could benefit from considering spectral decomposition.

\ack

Luciano da F. Costa is grateful to FAPESP (2005/00587-5) and CNPq
(308231/03-1) for financial sponsorship.

\bibliographystyle{unsrt}
\bibliography{spec}

\begin{thebibliography}{10}

\bibitem{Tutte:2001}
W.~T. Tutte.
\newblock {\em Graph Theory}.
\newblock Cambridge University Press, 2001.

\bibitem{Albert_Barab:2002}
R.~Albert and A.~L. Barab\'asi.
\newblock Statistical mechanics of complex network.
\newblock {\em Rev. Mod. Phys.}, 74:47--97, 2002.

\bibitem{Newman:2003}
M.~E.~J. Newman.
\newblock The structure and function of complex networks.
\newblock {\em SIAM Review}, 45:167--256, 2003.
\newblock cond-mat/0303516.

\bibitem{Boccaletti_etal:2006}
S.~Boccaletti, V.~Latora, Y.~Moreno, M.~Chavez, and D.-U. Hwang.
\newblock Complex networks: Structure and dynamics.
\newblock {\em Physics Reports}, 2006.

\bibitem{Costa_surv:2006}
L.~da~F.~Costa, F.~A. Rodrigues, G.~Travieso, and P.~Villas Boas.
\newblock Characterization of complex networks: {A} survey of measurements.
\newblock {\em Advances in Physics}, 56:167--242, 2007.

\bibitem{Biggs:1993}
N.~Biggs.
\newblock {\em Algebraic graph theory}.
\newblock Cambridge University Press, 1993.

\bibitem{Chung:1997}
F.~R. Chung.
\newblock {\em Spectral Graph Theory}.
\newblock American Mathematical Society, 1997.

\bibitem{Cvetkovic:1997}
Dragos Cvetokovic, Peter Rowlinson, and Slobodan Simic.
\newblock {\em Eigenspaces of Graphs}.
\newblock Cambridge University Press, 1997.

\bibitem{Newman:2006}
M.~E.~J. Newman.
\newblock Finding community structure in networks using the eigenvectors of
  matrices.
\newblock {\em Physical Review E}, 74:036104, 2006.

\bibitem{latora05:_vulner}
Vito Latora and Massimo Marchiori.
\newblock Vulnerability and protection of infrastructure networks.
\newblock {\em Phys. Rev. E}, 71(1):015103, Jan 2005.

\bibitem{holme02:_attac}
Petter Holme, Beom~Jun Kim, Chang~No Yoon, and Seung~Kee Han.
\newblock Attack vulnerability of complex networks.
\newblock {\em Phys. Rev. E}, 65(5):056109, May 2002.

\bibitem{He20092243}
Shan He, Sheng Li, and Hongru Ma.
\newblock Effect of edge removal on topological and functional robustness of
  complex networks.
\newblock {\em Physica A: Statistical Mechanics and its Applications},
  388(11):2243--2253, 2009.

\bibitem{martin06:_random}
S.~Martin, R.D. Carr, and J.-L. Faulon.
\newblock Random removal of edges from scale free graphs.
\newblock {\em Physica A: Statistical and Theoretical Physics},
  371(2):870--876, 2006.

\bibitem{motter02:_range}
Adilson~E. Motter, Takashi Nishikawa, and Ying-Cheng Lai.
\newblock Range-based attack on links in scale-free networks: Are long-range
  links responsible for the small-world phenomenon?
\newblock {\em Phys. Rev. E}, 66(6):065103, Dec 2002.

\bibitem{wang08:_robus}
Yubo Wang, Shi Xiao, Gaoxi Xiao, Xiuju Fu, and Tee~Hiang Cheng.
\newblock Robustness of complex communication networks under link attacks.
\newblock In {\em Proceedings of the 2008 International Conference on Advanced
  Infocomm Technology}, ICAIT '08, pages 61:1--61:7, New York, NY, USA, 2008.
  ACM.

\bibitem{gao06:_between}
Liang Gao, Menghui Li, Jinshan Wu, and Zengru Di.
\newblock Betweenness-based attacks on nodes and edges of food webs.
\newblock {\em Dynamics of Continuous, Discrete and Impulsive Systems},
  B13:421--428, 2006.

\bibitem{estrada08:_graph}
Ernesto Estrada.
\newblock Graph spectra and the structure of complex networks.
\newblock Technical Report~26, University of Strathclyde, 2008.

\bibitem{ma10:_eigen}
Xiaoke Ma, Lin Gao, and Xuerong Yong.
\newblock Eigenspaces of networks reveal the overlapping and hierarchical
  community structure more precisely.
\newblock {\em Journal of Statistical Mechanics: Theory and Experiment},
  2010(08):P08012, 2010.

\bibitem{qin09:_commun_findin_scale_free_networ}
Sen Qin and Guanzhong Dai.
\newblock Community finding of scale-free network: Algorithm and evaluation
  criterion.
\newblock In Djamel Zighed, Shusaku Tsumoto, Zbigniew Ras, and Hakim Hacid,
  editors, {\em Mining Complex Data}, volume 165 of {\em Studies in
  Computational Intelligence}, pages 223--242. Springer Berlin / Heidelberg,
  2009.

\bibitem{yazdani10}
A.~Yazdani and P.~Jeffrey.
\newblock A note on measurement of network vulnerability under random and
  intentional attacks.
\newblock arXiv:1006.2791, 2010.

\bibitem{green09:_small_scotl}
Darren~Michael Green, Alison Gregory, and Lorna~Ann Munro.
\newblock Small- and large-scale network structure of live fish movements in
  scotland.
\newblock {\em Preventive Veterinary Medicine}, 91(2-4):261 -- 269, 2009.

\bibitem{jamakovic07}
A.~Jamakovic and S.~Uhlig.
\newblock On the relationship between the algebraic connectivity and graph's
  robustness to node and link failures.
\newblock In {\em Next Generation Internet Networks, 3rd EuroNGI Conference
  on}, pages 96--102, may 2007.

\bibitem{PhysRevE.81.016101}
D.~Liu, H.~Wang, and P.~Van~Mieghem.
\newblock Spectral perturbation and reconstructability of complex networks.
\newblock {\em Phys. Rev. E}, 81(1):016101, Jan 2010.

\bibitem{HolmeKim:2002}
Petter Holme and Beom~Jun Kim.
\newblock Growing scale-free networks with tunable clustering.
\newblock {\em Physical Review E}, 65:026107, 2002.

\bibitem{pajekdata}
Vladimir Batagelj and Andrej Mrvar.
\newblock Pajek datasets, 2006.

\bibitem{kaiser04:_edge}
Marcus Kaiser and Claus~C. Hilgetag.
\newblock Edge vulnerability in neural and metabolic networks.
\newblock {\em Biological Cybernetics}, 90:311--317, 2004.

\end{thebibliography}
\end{document}